\documentstyle[psfig,nicV]{article}

\newcommand{\ctanb}{$^{13}$C($\alpha$,n)$^{16}$O~}

\newcommand{\neanb}{$^{22}$Ne($\alpha$,n)$^{25}$Mg~}

\newcommand{\msb}{$M_{\odot}$~}

\newcommand{\ms}{$M_{\odot}$}

\begin{document}
%
%
\heading{%
Galactic chemical evolution of Ba-peak elements
%
}
\par\medskip\noindent
%
\author{Claudia Travaglio$^{1}$, Daniele Galli$^{2}$, Roberto
Gallino$^{3}$, Maurizio Busso$^{4}$, Federico Ferrini$^{5}$,
Oscar Straniero$^{6}$}
\address{Dip. di Astronomia e Scienze dello Spazio, 
        Largo Fermi 5, I-50125 Firenze, Italy}
\address{Osservatorio Astrofisico di Arcetri, Largo Fermi 5, 
        I-50125 Firenze, Italy}
\address{Dip. di Fisica Generale, Universit\`a di Torino, 
         Via P.Giuria 1, I-10125 Torino, Italy}
\address{Osservatorio Astronomico di Torino, Strada Osservatorio 20,
        I-10025 Torino, Italy} 
\address{Dip. di Scienze, Universit\`a di Pisa, Piazza Torricelli 
        2, I-56100 Pisa, Italy}
\address{Osservatorio di Collurania, I-64100 Teramo, Italy}
%
\begin{abstract}
The chemical evolution of the Galaxy is followed for the elements 
affected by neutron capture, in particular for those in the atomic 
number range 56 to 63 (Ba, La, Ce, Pr, Nd, Sm and Eu). 
Neutrons by the major $^{13}$C source, released in radiative
conditions in the interpulse periods of TP-AGB stars, give rise to an
efficient $s$-processing, making low mass AGB the major
contributors to the chemical evolution of heavy elements. 
The $s$-process scenario,
characterized by the combined operation of the two neutron sources \ctanb 
and \neanb, is analyzed using AGB stellar evolutionary calculations with
the FRANEC Code (FRASCATI Raphson-Newton Evolutionary Code), and are
applied over a wide range of stellar masses and metallicities. 
The presence of $r$-process elements in low metallicity stars 
is indicative of a prompt enrichment of the Galaxy by early generation
of stars, and low mass SNII appear to be good candidates for
{\em primary} production of $r$-nuclei. 
The chemical evolution model used here is organized over three-zone,
halo, thick and thin disk. A comparison between model abundance
predictions of the $r$- and $s$-process elements observed in unevolved
halo and disk stars confirms the overall consistency of the theoretical
framework and reveals a number of striking features deserving a careful
analysis.

\end{abstract}

\section{$s$-process from Asymptotic Giant Branch Stars}
Since the pioneering work on stellar nucleosynthesis by Burbidge et al.
~(1957), the synthesis of nuclei heavier than iron has been recognized 
to be dominated by neutron capture processes, both {\em slow} (the
$s$-process), and {\em rapid} (the $r$-process). 
For the $s$-process, the abundance distribution in the solar system is 
currently considered as the superposition of two components, the $weak$ 
and the $main$ component. The weak component, responsible for the 
$s$-process nuclides up to $A \approx 90$, is due to neutron captures
occurring in massive stars by the activation of the \neanb reaction.
The main component, feeding the heavier $s$-process nuclides, 
originates in low mass AGB stars during
the recurrent thermal instabilities in the He shell. 
Many theoretical and observational works converge on the idea that
neutrons are released in radiative conditions in the interpulse period
via the \ctanb reaction. A tiny $^{13}$C-pocket is assumed to develop
as a consequence of the penetration of a small amount of protons from 
the envelope in the $^{12}$C-rich intershell.
The $s$-process yields adopted here have been obtained performing
post-process calculations starting from stellar evolutionary models
obtained with the FRANEC code (Straniero et al.~1997; Gallino et
al.~1998). In these models, the third dredge-up mechanism,
mixing with the envelope newly synthesized $^{12}$C and 
$s$-process elements, is self-consistently found after a limited 
number of thermal pulses. Concerning the dependence of the 
results on the Galactic chemical evolution of the $s$-elements on 
the choice of the amount and profile of the $^{13}$C-pocket and on 
the dredged-up mass in the envelope we refer to the discussion in 
Travaglio et al.~(1998).
Since the $^{13}$C-pocket is of primary origin, 
the $s$-process distribution is strongly dependent on metallicity 
(Clayton~1988). Typical production factors (with respect to solar) of 
elements belonging to the three major $s$-process peaks are shown in 
Fig. 1 as a function of metallicity. The production factor of Eu, an 
element mostly fed by the $r$-process, is also shown for comparison. 
For more details see Gallino et al.~(1999).
\begin{figure}
\centerline
{\vbox{\psfig{figure=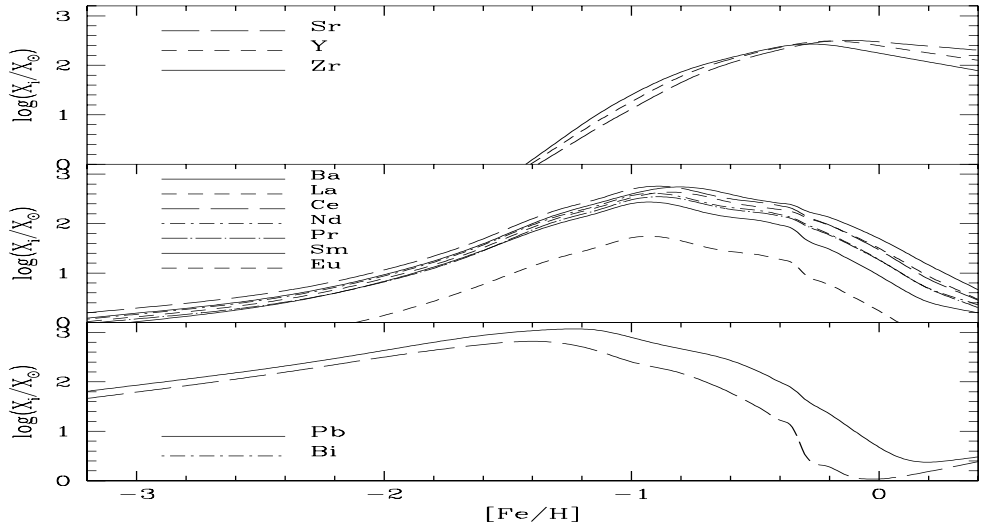,height=8.0cm,width=15cm}}}
\caption[]{\small Production factors as a function of metallicity
of selected elements in the He shell material cumulatively 
mixed with the surface of a $M = 2$ \msb AGB star.}
\end{figure}

\section{Chemical evolution model}
The chemical evolution model adopted for this work is described in
Ferrini et al.~(1992). It is based on the interconnected evolution
of three zones: halo, thick disk and thin disk, whose relative 
composition in stars, gas phases, and stellar remnants, is followed 
during the Galactic age. The thin disk is divided into concentric 
annuli. Here we consider the evolution of the solar annulus,
located 8.5 kpc from the Galactic center. The Star Formation Rate 
(SFR) is obtained as outcome of self-regulating processes occurring 
in the molecular gas phase, either
spontaneous or stimulated by the presence of massive stars. In Fig. 2 
the SFR is plotted as a function of [Fe/H] during the evolution of the
Galaxy. 

\begin{figure}
\begin{minipage}{5.6cm}
{\psfig{figure=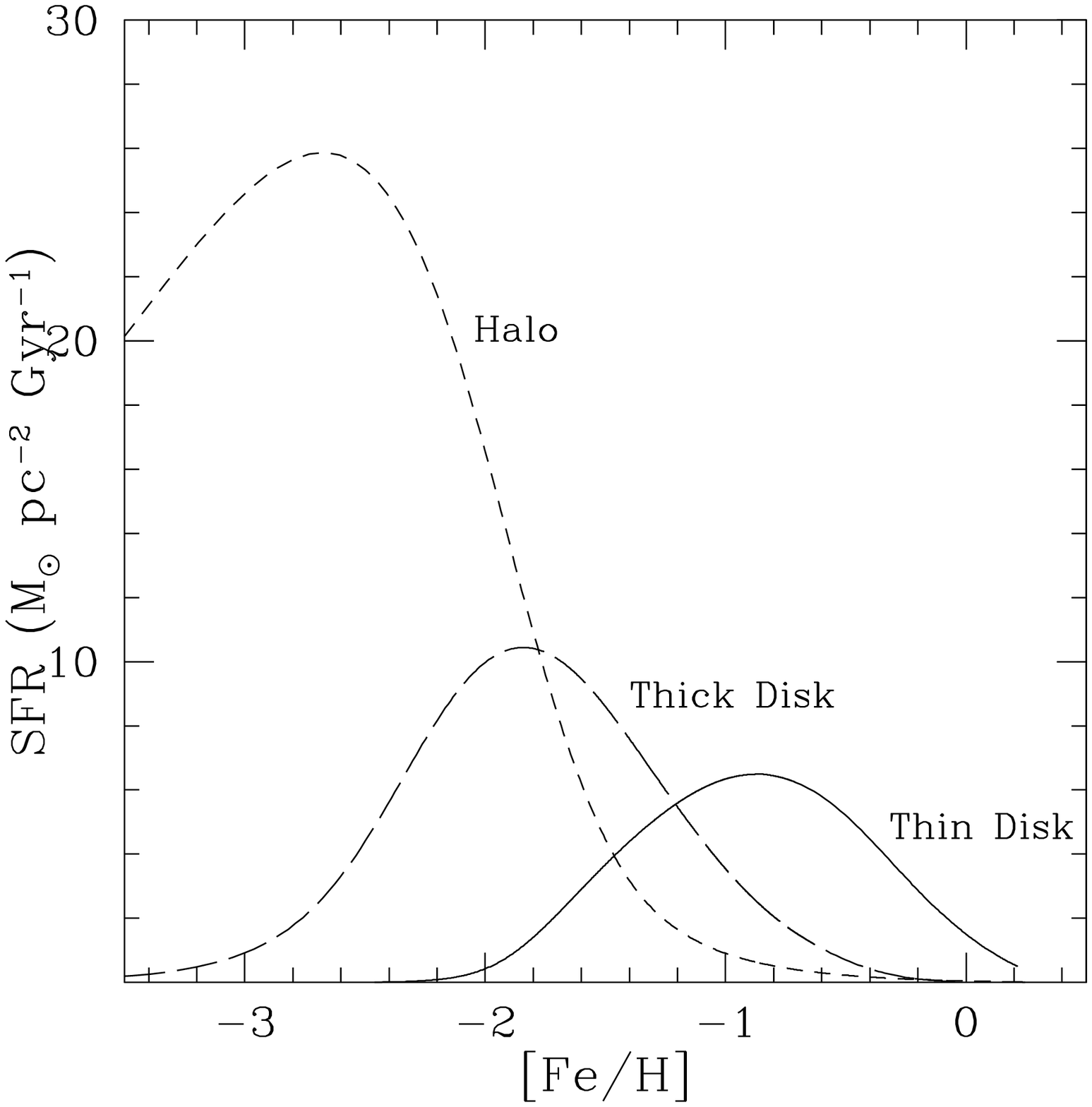,height=6.5cm,width=6.0cm}}
\caption[]{\small Star Formation Rate, according to our model prediction,
as a function of [Fe/H].}
\end{minipage} 
\begin{minipage}{6.4cm}
{\psfig{figure=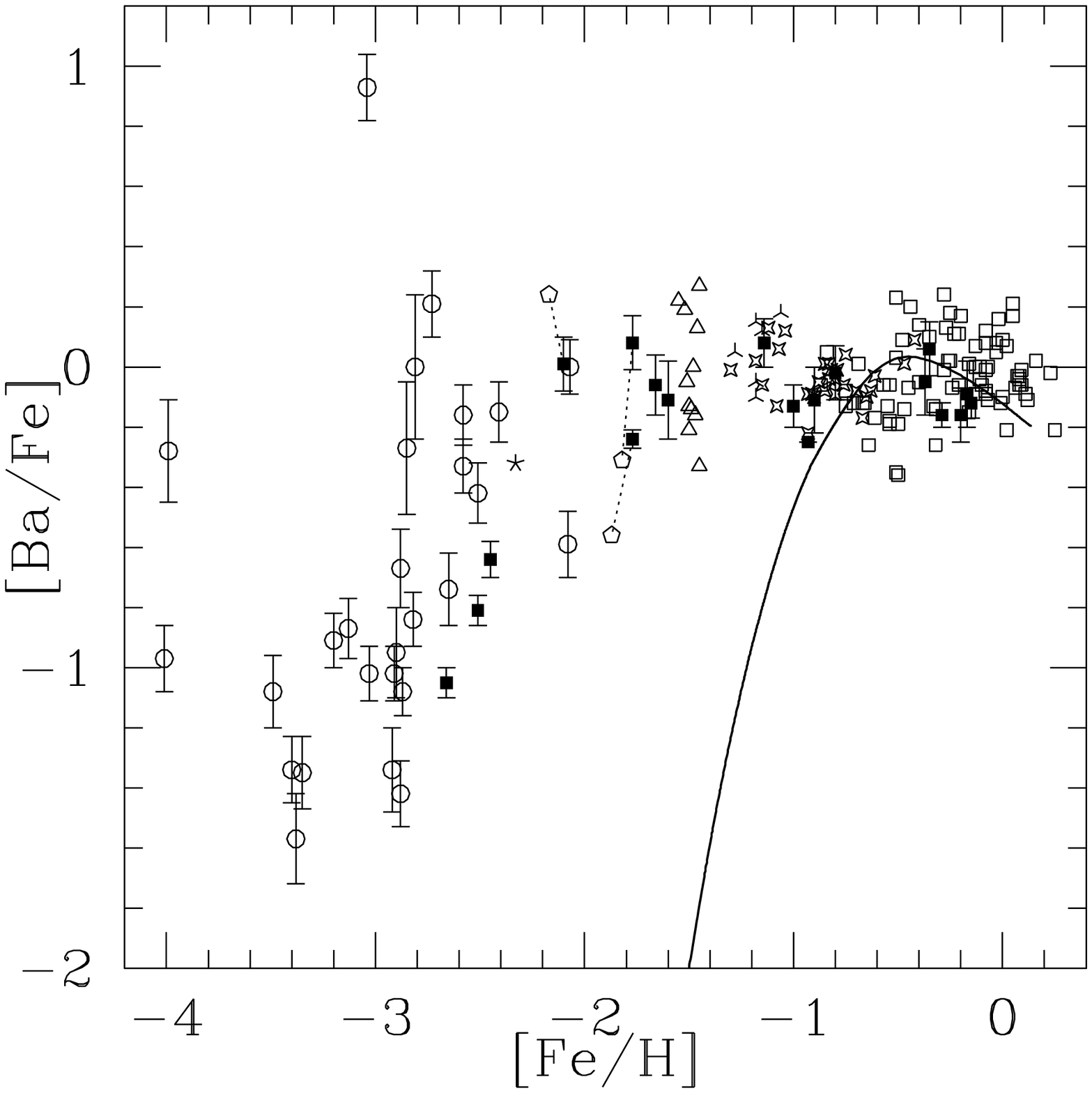,height=4.5cm,width=6.0cm}}
\caption[]{\small Evolution of the Ba $s$-fraction vs. [Fe/H]
according to our standard model. {\em Solid} line refers to thin-disk; 
different symbols are used for the available observational data.
Thin dotted lines connects the same stars with abundance determinations
by different authors.}
\end{minipage}
\end{figure}

\section{Galactic evolution of Ba from AGB stars}
The chemical evolution of Ba, taken as
representative of the heavy $s$-process elements, is plotted 
in Fig. 3 and is compared with spectroscopic observations of unevolved 
stars of different metallicities. 
The lower and upper limit of AGB stellar masses were 
taken to be 2 and 4 \ms, respectively, similar to the observed range 
of chemically peculiar AGB stars.
At $t = t_\odot$ we obtain a Ba $s$-process contribution
to solar of 80$\%$. From Fig. 3 one can see that the $s$-process 
contribution
begins to dominate the galactic evolution at [Fe/H] $\simeq - 1.5$,
whereas at lower [Fe/H] the contribution of low-mass AGB stars is 
by far too low. This is essentially due to the
strong dependence of the $s$-process yields on metallicity.
To check the sensitivity of this result, Travaglio et al.~(1998)
also considered the effect of adding the $s$-process contribution
by intermediate mass AGB stars in the range $4 - 8$ \ms, starting from 
FRANEC evolutionary models of a 5 \msb and a 7 \msb (Vaglio et 
al.~1999). However, the efficiency of these more massive stars in 
producing Ba is too low to contribute significantly to the 
chemical enrichment of the Galaxy.

\section{$r$-process treatment in our model}
As first stressed by Truran~(1981), the heavy element abundance 
patterns in very metal-poor stars are compatible with  
an $r$-process origin. This point was been recently sustained by 
new HST observations of low metallicity unevolved stars, reported by 
Sneden et al.~(1998).
From the theoretical point of view, despite a large number of recent 
works, the astrophysical site of the $r$-process is still uncertain 
(Baron et al.~1998; Wheeler at al.~1998).
In order to quantify the $r$-contribution, we treated 
the $r$-process as a typical {\em primary} process originating from 
low-mass Type II SNe ($M$ = 8 - 10 \ms), in agreement with recent 
theoretical predictions from Wheeler et al.~(1998). Then, at $t = 
t_\odot$ we derived the {\em $r$-residuals}, after subtracting from 
the solar abundances the predicted $s$-fractions.

\section{Galactic evolution of Ba and Eu}
In Fig. 4 the predicted Galactic evolutionary trends for the $s + r$
contributions to Ba and Eu are shown, as compared with the available 
spectroscopic data for unevolved halo and disk stars. 
These plots make clear that a delay in the $r$-process production with
respect to other heavy elements from SNII (i.e. Fe) is needed in order 
to match the spectroscopic data at [Fe/H]$<$-2.
\begin{figure}
\centerline
{\vbox{\psfig{figure=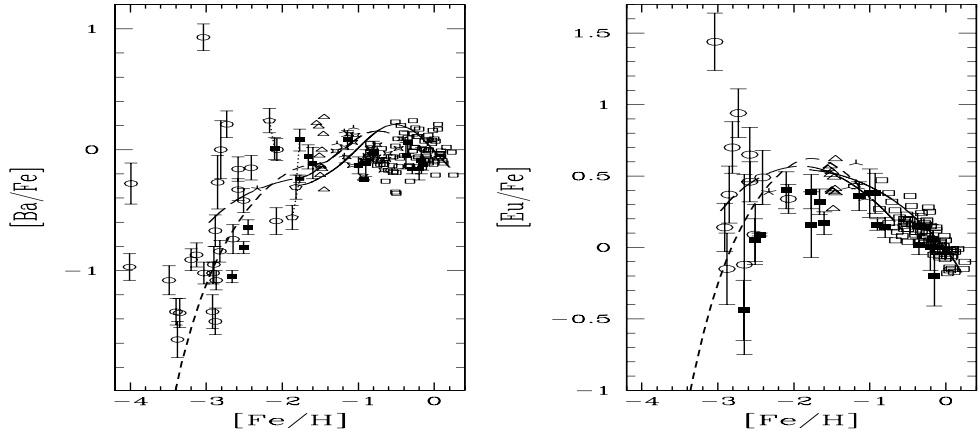,height=6.5cm,width=14.5cm}}}
\caption[]{\small Evolution of [Ba/Fe] and [Eu/Fe] vs. [Fe/H] according to
our model predictions for halo ({\em short-dashed}), thick disk ({\em
long-dashed}) and thin disk ({\em solid}), compared with the
available spectroscopic observations.}
\end{figure}
Since Eu is mostly made by $r$-process nucleosynthesis (94$\%$ at $t =
t_\odot$), the predicted [Eu/Fe] trend vs. [Fe/H], compared with
observations, is a good test of our assumption to quantify the 
$r$-process fraction in the Galaxy. The declining in [Ba/Fe] and in 
[Eu/Fe] at [Fe/H] $\simeq$ -2.5 can be explained by the finite lifetime 
of the low-mass SNII, hence discriminating the production 
of the $r$-process from the Fe production in more massive SNe. 
The large observational scatter for the stars at the lowest 
metallicities can be essentially attributed to chemical heterogeneity 
in the Galactic halo.
In conclusion, the Galactic barium enrichment can be explained through 
the interplay of a complex $s$-process mechanism taking place in low 
mass AGB stars at various metallicities, and of a primary-like
$r$-process occurring in the lower mass range of SNIIe.

\vfill
\end{document}